\documentstyle[11pt]{article}
\input psfig
\long\def\comment#1{}
\begin{document}
\title{A Three-Stage Quantum Cryptography Protocol}
\author{Subhash Kak\\
Department of Electrical \& Computer Engineering\\
Louisiana State University,
Baton Rouge, LA 70803, USA}
\maketitle

\begin{abstract}
We present a three-stage quantum cryptographic protocol based on 
public key cryptography in which each party uses its own secret key.
Unlike the BB84 protocol, where the qubits are transmitted
in only one direction and classical information exchanged thereafter, the communication
in the proposed protocol remains quantum in each stage.
A related system of key distribution is also described.

\end{abstract}

\thispagestyle{empty}

\subsection*{Introduction}

This paper presents a quantum protocol based on public
key cryptogrpahy for secure transmission of data over
a public channel.
The security of the protocol derives from the fact that Alice and Bob each use
secret keys in the multiple exchange of the qubit.

Unlike the BB84 protocol [1] and its many variants (e.g. [2]-[4]), 
where the qubits are transmitted
in only one direction and classical information exchanged thereafter, the communication
in the proposed protocol remains quantum in each stage.
In the BB84 protocol, each transmitted qubit is in one of four different states;
in the proposed protocol, the transmitted qubit can be in any arbitrary state.

\subsection*{The Protocol}

Consider the arrangement of Figure 1 to transfer state $X$ from Alice to Bob.
The state $X$ is one of two orthogonal states, such as $0\rangle$ and $|1\rangle$,
or $\frac{1}{\sqrt2} (| 0 \rangle +  | 1\rangle )$
and $\frac{1}{\sqrt2} (| 0 \rangle -  | 1\rangle )$, or
$\alpha |0\rangle + \beta |1\rangle$ and 
$\beta |0\rangle - \alpha |1\rangle$.
The orthogonal states of $X$ represent $0$
and $1$ by prior mutual agreement of the parties, and this is the data
or the cryptographic key being transmitted over the public channel.

Alice and Bob apply secret transformations $U_A$ and $U_B$ which are commutative,
i.e., $U_A U_B = U_B U_A$.
An example of this would be $U_A = R(\theta)$ and $U_B = R (\phi)$, each of which
is the rotation operator:

\vspace{0.2in}
$R(\theta) =                        \left[ \begin{array}{cc}
cos \theta  & - sin \theta  \\
sin \theta  & cos \theta  \\
\end{array} \right]$

\vspace{0.2in}

\begin{figure}
\hspace*{0.2in}\centering{
\psfig{file=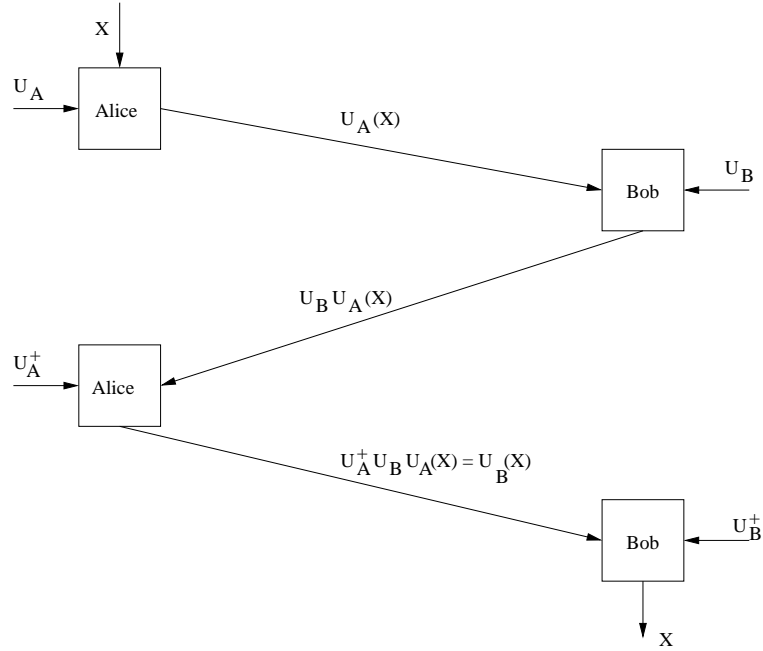,width=10cm}}
\caption{Three-stage protocol for quantum cryptography where $U_A U_B = U_B U_A$}
\end{figure}

\vspace{0.2in}

\noindent
The sequence of operations in the protocol is as follows:

\begin{enumerate}

\item Alice applies the transformation $U_A$ on $X$ and sends the qubit to Bob.

\item Bob applies $U_B$ on the received qubit $U_A (X)$ and sends it
back to Alice.

\item
Alice applies $U_A^\dagger$ on the received qubit, converting it to 
$U_B (X)$, and forwards it to Bob. 

\item
Bob applies $U_B^\dagger$ on the qubit, converting it to $X$.

\end{enumerate}

At the end of the sequence, the state $X$, which was
chosen by Alice and transmitted
over a public channel, has reached Bob.

Eve, the eavesdropper, cannot obtain any information by intercepting the 
transmitted qubits, although she could disrupt the exchange by replacing the
transmitted qubits by her own. This can be detected by 

\begin{itemize}
\item appending parity bits, and/or
\item appending previously chosen bit sequences, which could be the destination and
sending addresses or their hashed values, or some other mutually agreed sequence.
\end{itemize}

Since the $U$ transformations can be changed as frequently as one pleases, Eve
cannot obtain any statistical clues to their nature by intercepting the qubits.

\subsection*{Key distribution protocol}

A related key distribution 
protocol is given in Figure 2.
Unlike the previous case, $X$ is a fixed public state (say $|0\rangle$ or
$\frac{1}{\sqrt2} (|0\rangle + |1\rangle)$). 
The objective is to generate a key that is a function of the
transformations involved, which is not chosen in advance by either party.
The protocol consists of two stages:

\begin{figure}
\hspace*{0.2in}\centering{
\psfig{file=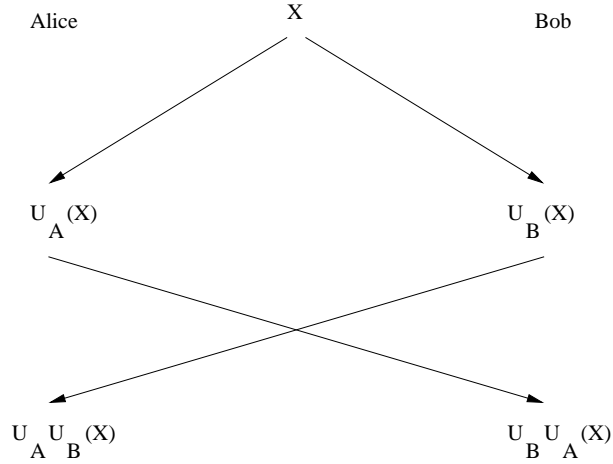,width=8cm}}
\caption{Key distribution protocol, where $U_A U_B = U_B U_A$.}
\end{figure}

\begin{enumerate}

\item Alice and Bob use secret transformations, $U_A$ and $U_B$, on the known
state $X$, and exchange these qubits.
\item 
They again apply the same transformations on the received qubits, thereby each 
getting $U_A U_B (X)$, since $U_A U_B = U_B U_A$. It is assumed that neither
Alice or Bob will measure the received qubits, and will use them as the input
to a quantum register.
\end{enumerate}

In a variant of this scheme, two copies of the unknown state $X$ may be supplied
to Alice and Bob by a key registration authority.

\subsection*{Conclusion}
The three-stage protocol provides perfect security in the exchange of data over a public 
channel under the assumptions that a separate classical protocol ensures the
identity of the two parties, and errors (deliberate or random) are detected
by means of parity check and confirming that
a known bit sequence that was appended to the bits has arrived correctly.

Since the proposed protocol does not use
classical communication, it is immune to
the man-in-the-middle attack on the classical 
communication channel which BB84 type quantum cryptography protocols
suffers from [5].
On the other hand, implementation of this protocol may be harder because
the qubits get exchanged multiple times. 
\section*{References}
\begin{description}

\item
[1]
M.A. Nielsen and I.L. Chuang, {\it Quantum Computation and Quantum Information}.
Cambridge University Press, 2000.

\item
[2] 
A.K. Ekert, ``Quantum cryptography based on Bell's theorem.'' 
Phys. Rev. Lett., {\bf 67,} 661-663 (1991).
\item
[3] S. Kak, ``Quantum key distribution using three basis states.''
Pramana, {\bf 54,} 709-713 (2000); also quant-ph/9902038.
\item
[4]
A. Poppe {\it et al}, ``Practical quantum key distribution with polarization entangled
photons.'' quant-ph/0404115.
\item
[5] K. Svozil, ``The interlock protocol cannot save quantum cryptography from
man-in-the-middle attacks.'' quant-ph/0501062.


\end{description}
 
\end{document}